\documentclass[sigconf,authorversion]{acmart}

\usepackage{float}
\usepackage{tabularx}
\usepackage{amsmath}
\usepackage{graphicx}
\usepackage{booktabs}

\AtBeginDocument{%
}

\copyrightyear{2026}
\acmYear{2026}
\setcopyright{cc}
\setcctype{by}
\acmConference[L@S '26]{Proceedings of the Thirteenth ACM Conference on Learning @ Scale}{June 29-July 03, 2026}{Seoul, Republic of Korea}
\acmBooktitle{Proceedings of the Thirteenth ACM Conference on Learning @ Scale (L@S '26), June 29-July 03, 2026, Seoul, Republic of Korea}
\acmDOI{10.1145/3774398.3811624}
\acmISBN{979-8-4007-2293-6/2026/06}

\begin{document}
\title[Understanding Student Effort Using Response-Time Propensities]{Understanding Student Effort Using Response-Time Propensities During Problem Solving}

\author{Conrad Borchers}
\orcid{0000-0003-3437-8979}
\affiliation{
    \institution{Carnegie Mellon University}
    \city{Pittsburgh}
    \state{PA}
    \country{USA}
}
\email{cborcher@cs.cmu.edu}

\author{Lijin Zhang}
\orcid{0000-0002-4222-8850}
\affiliation{
    \institution{Stanford University}
    \city{Stanford}
    \state{CA}
    \country{USA}
}
\email{lijinzhang@stanford.edu}

\author{Kexin Yang}
\orcid{0000-0002-8132-0166}
\affiliation{
    \institution{Carnegie Mellon University}
    \city{Pittsburgh}
    \state{PA}
    \country{USA}
}
\email{kexiny@cs.cmu.edu}

\author{Tomohiro Nagashima}
\orcid{0000-0003-2489-5016}
\affiliation{
    \institution{Saarland University}
    \city{Saarbrücken}
    \country{Germany}
}
\email{nagashima@cs.uni-saarland.de}

\author{Benjamin W. Domingue}
\orcid{0000-0002-3894-9049}
\affiliation{
    \institution{Stanford University}
    \city{Stanford}
    \state{CA}
    \country{USA}
}
\email{bdomingue@stanford.edu}

\renewcommand{\shortauthors}{Conrad Borchers, Lijin Zhang, Kexin Yang, Tomohiro Nagashima, \& Benjamin W. Domingue}

\begin{abstract}
Adaptive learning systems can produce substantial learning gains, yet many students engage for too brief or too superficial a period to benefit. A central obstacle is measuring effort. Effort during multi-step problem solving is rarely directly observed, and common log-based proxies, such as time on task, cannot distinguish between a student working carefully and a student encountering a harder problem. We examine step-to-step response time as a scalable effort signal by modeling trait-like differences in students' typical response timing during tutoring (while adjusting for skill difficulty). Using step-level logs from eight classroom deployments of algebra tutoring systems (2020 to 2023) across six U.S. schools (794 students), we estimate student- and knowledge-component-level propensities using hierarchical models and relate them to learning efficiency, defined as performance improvement per completed solution step. Response-time propensities show moderate to strong stability within students, supporting their use as an individual differences measure beyond correctness. At the same time, their relationship to learning is not uniform but conditional on the learner and context. Slower propensities predict greater learning efficiency for higher-proficiency students, consistent with constructive processing, whereas for lower-proficiency students, slower propensities are weakly related or even negative, consistent with unproductive struggle or idling. These associations are strongest early in practice sequences and attenuate later in the class period, highlighting an actionable window for detecting emerging disengagement and low persistence. Overall, response-time propensities provide a practical way to incorporate temporal process data into learner models and to target adaptive supports when effort is most diagnostic.
\end{abstract}
\begin{CCSXML}

<concept_id>10003120.10003121.10011748</concept_id>
<concept_desc>Human-centered computing~Empirical studies in HCI</concept_desc>
<concept_significance>300</concept_significance>

<concept_id>10010405.10010455.10010459</concept_id>
<concept_desc>Applied computing~Psychology</concept_desc>
<concept_significance>300</concept_significance>

<concept_id>10010405.10010489</concept_id>
<concept_desc>Applied computing~Education</concept_desc>
<concept_significance>500</concept_significance>

\end{CCSXML}

\ccsdesc[300]{Human-centered computing~Empirical studies in HCI}
\ccsdesc[300]{Applied computing~Psychology}
\ccsdesc[500]{Applied computing~Education}
\keywords{intelligent tutoring systems, latent trait modeling, effort regulation}
\maketitle

\section{Introduction}

Adaptive learning technologies and intelligent tutoring systems (ITSs) have repeatedly shown strong potential to improve student learning by supporting active, step-by-step problem solving at scale \cite{koedinger2015learning}. However, evidence from large-scale classroom deployments paints a more sobering picture. In many real-world settings, only a small fraction of students engage with these systems long enough, or deeply enough, to experience their intended learning benefits. Recent analyses suggest that only a small proportion of students (sometimes as low as 5\%) persist at levels sufficient to meaningfully benefit from adaptive learning technologies \cite{eames2026computer,holt20245}. These results point to substantial gaps in student engagement and persistence, and they highlight a core challenge for learning at scale: \textbf{systems designed to be effective do not guarantee effective use by learners}. Accordingly, our field needs better methods to \textit{measure and explain} between-student differences in effort. The present study addresses this gap by modeling step-level response time as a difficulty-adjusted, trait-like effort signal and examining when it predicts learning.

We use \emph{effort} to mean behavioral engagement during practice: how students allocate time and attention across steps and respond to task demands. \emph{Persistence} refers to \textit{sustained effort} over time (e.g., completing practice sessions) even when encountering challenges, errors, or difficulties.
Both effort regulation and persistence as adaptation to errors are central components of self-regulated learning \cite{locke2019development,Panadero_2017,Zimmerman_2000}.
Understanding effort and persistence during learning is, therefore, a critical problem.
In ITS research, they matter not only because they influence whether students complete practice, but because they shape how students interact with feedback, hints, and problem-solving opportunities.
Yet most learner modeling work has focused on modeling \textit{what students know} by tracking accuracy and performance over time \cite{koedinger2023astonishing} or on detecting when students momentarily disengage \cite{baker2007gaming}.
These models are highly effective for estimating knowledge and instructional effectiveness \cite{borchers2024combining,koedinger2012knowledge}, but they provide only limited insights into how students regulate effort while solving problems. Large-scale studies using these same systems have documented dramatic differences in student persistence and productive use of time \cite{eames2026computer,holt20245}.
Recent work has begun to address this gap by examining stable individual differences in behaviors, such as how promptly students begin work or how much of a class session they spend engaged in practice \cite{gurung2025starting}. However, these measures largely operate at the session level and do not capture how effort unfolds within multi-step problem solving.

One promising signal of effort during problem solving is \textit{response time} between steps. Step-level response times are available at scale in most tutoring systems and reflect the temporal footprint of students' cognitive and behavioral engagement. Extremely fast responses may indicate shallow processing, guessing, or disengaged behavior, while longer response times can reflect deliberation, monitoring, or productive struggle \cite{chan2022slow,lee2025unpacking}. Alternatively, faster response times might also reflect improved task fluency \cite{wang2020using}. Prior work has shown that response times can be diagnostic for specific instructional events at scale, such as whether students read and use hints productively \cite{gurung2021examining}. At the same time, response time is a deeply ambiguous measure \cite{domingue2022speed,kovanovic2015penetrating}. Prior work on performance assessment shows that the relationship between response time and accuracy varies across tasks \cite{domingue2022speed}.
Longer times may reflect effortful reasoning, but they may also reflect confusion, off-task behavior, or simply harder content. As a result, raw response time is difficult to interpret across students and problems, limiting its usefulness as a general measure of effort.

This ambiguity points to a key limitation in existing work. Response time has rarely been modeled in a way that separates \textit{stable between-student differences} (i.e., traits) from \textit{contextual variation} due to content, task structure, or practice progression. Related work has begun to address this issue by using trait-style models to estimate students' propensities for behaviors such as gaming the system while adjusting for task difficulty \cite{huang2023using}. However, comparable trait-like models of \textbf{constructive effort during problem solving} remain underdeveloped. This gap is particularly important because theory and prior evidence suggest that the time spent on individual cognitive operations depends on both students' prior knowledge and content complexity \cite{kalyuga2009expertise}. For students with higher prior knowledge, spending more time on a step may reflect constructive processing such as self-explanation. For lower-knowledge students, longer delays may instead signal idling or unproductive struggle. Without models that explicitly account for these differences, response time risks being ineffective for predicting learning at scale.

In this paper, we address this gap by introducing \textbf{response-time propensities}, a log-derived, difficulty-adjusted, trait-like measure of students' typical step-to-step response timing in intelligent tutoring systems. Using step-level data from eight classroom deployments of algebra tutoring systems conducted between 2020 and 2023, spanning six U.S. schools and 794 students, we estimate student- and knowledge-component-level response-time propensities using hierarchical models. We then examine how these propensities relate to learning process outcomes, operationalized as learning efficiency or performance improvement per completed solution step. Critically, we test when and for whom response time is informative by examining moderation by prior proficiency and by early versus late segments of a class period. We contribute novel empirical evidence that response-time propensities are sufficiently stable to support an individual-differences interpretation; that their relationship to learning is moderated by prior proficiency and position within the session; and that they offer a scalable way to incorporate temporal-process data into learner models when interpreted in context.

\subsection{The Present Study}
\label{sec:present-study}

The present study investigates the stability and predictive value of step-level response time during multi-step problem solving in ITSs. Raw response time is ambiguous (e.g., fast steps can reflect either efficiency or gaming; slow steps can reflect deliberation or idling). We therefore estimate difficulty-adjusted, trait-like propensities and ask when they relate to learning. Unlike correctness-based measures alone, response times can reveal patterns of productive struggle, disengagement, or strategic behavior. Our approach estimates both student-level and knowledge-component-level propensities to capture individual differences in typical pacing and to separate them from task difficulty.

Specifically, we address three research questions:

\noindent\textbf{RQ1}: How stable are student-level response-time propensities during multi-step problem solving?

\noindent\textbf{RQ2}: To what extent do student-level and skill-level response-time parameters predict learning efficiency in intelligent tutoring systems?

\noindent\textbf{RQ3}: How do contextual factors, such as early versus late problem-solving opportunities and student prior proficiency, influence the relationship between response time and learning outcomes?

We analyze large-scale log data from multiple ITS deployments, applying hierarchical models to extract student- and skill-level parameters. This analysis allows us to quantify the stability of response-time propensities, understand situational influences on engagement, and evaluate their relationship to learning efficiency. The study provides a framework for integrating response-time measures into adaptive systems to better support student learning at scale.

\section{Background}

In this section, we review prior research on effort regulation, log-based measures of student effort such as response time, and trait-based modeling of learner behavior in adaptive learning systems. This review highlights both what is known about effort at scale and the gaps our work addresses.

\subsection{Effort Regulation}

Effort regulation refers to learners' capacity to initiate, sustain, and adapt effort in response to task demands and feedback. It is a central component of self-regulated learning (SRL) \cite{locke2019development,Panadero_2017,Zimmerman_2000}.
In problem-solving environments, especially those requiring multiple steps to solve a single problem, the effectiveness of instruction can depend on how students allocate time and attention across steps, respond to errors, and persist when progress is slow. For instance, students may be demotivated by failure during practice and subsequently shift to easier content \cite{vanacore2025downshifting}, limiting their ability to learn \cite{kalyuga2009expertise}. These processes unfold through repeated, fine-grained decisions that are directly observable in digital learning environments.

At scale, however, effort regulation is often approximated using coarse indicators such as total time on task or activity counts \cite{gurung2025starting,kovanovic2015penetrating}. Prior work shows that these proxies are limited, as they can conflate productive cognitive engagement with inactivity or off-task behavior and are sensitive to arbitrary preprocessing choices \cite{kovanovic2015does,tempelaar2020learning}. More recent studies, therefore, emphasize log-based measures that are closer to regulatory decisions, such as task initiation, early stopping, or persistence after failure \cite{gurung2025starting,vanacore2025downshifting}, which are reliable and predictive of achievement and test scores beyond conventional engagement measures in tutoring systems such as gaming the system \cite{gurung2025starting}. Still, most of this work focuses on whether students persist over time rather than on how effort is allocated during problem solving.

Response times during multi-step problem solving provide a promising, scalable signal of effort regulation. From an SRL perspective, longer response times can reflect deliberation, monitoring, or productive struggle \cite{chan2022slow, lee2025unpacking}. In contrast, very short times may indicate rapid guessing or disengaged behaviors such as gaming the system \cite{baker2008students,gurung2021examining}. At the same time, response times are strongly shaped by task difficulty and step structure, making raw timing measures difficult to interpret across items and learners without adjustment. This points to a key research gap in learning at scale: the need for difficulty-adjusted models that disentangle student-level effort propensities from task-level effects using response-time data. The present study addresses this gap by estimating student- and skill-level response-time parameters from large-scale ITS logs and evaluating how these latent effort propensities relate to persistence and learning efficiency over time. By grounding response-time measurement in latent-trait modeling, this work provides a principled approach for assessing effort regulation and supporting adaptive learning systems at scale.

\subsection{Effort Measurement}

In large-scale digital learning and computer-based assessment, researchers rarely directly observe effort. Instead, effort is typically operationalized as \emph{behavioral engagement}: the extent to which students actively interact with a task, persist, and respond to system prompts. This definition is intentionally pragmatic. It treats effort as something the platform can reliably measure in logs, rather than as an internal motivational state. Within this measurement perspective, response time is often interpreted as an indicator of students' \emph{willingness to engage} with a curricular sequence of problems over a prolonged period of time (as opposed to a direct measure of motivation) \cite{WiseKong2005}. Alongside response time, common log-based indicators include time on task, counts of interface actions (e.g., clicks or keystrokes), and participation or completion rates (Table~\ref{tab:effortmeasure}). These features are attractive because they are ubiquitous, inexpensive to collect, and comparable across large populations and deployments (though predictive models tend to be hard to generalize between platforms and populations \cite{baker2019challenges}).

\begin{table}[htpb]
\centering
\caption{Effort measurement in prior research.}
\label{tab:effortmeasure}
\small
\begin{tabularx}{\columnwidth}{l l X}
\toprule
\textbf{Citation} & \textbf{Measure} & \textbf{Operationalization} \\
\midrule
\cite{Li2022} & Time on task  & Total interaction duration. \\

\cite{Ivanova2020} & Number of actions & Clicks, double-clicks, key presses, drag/drop events. \\

\cite{IvanovaMichaelides2023} & Response time   &  Total time a student spends on an item. \\
\cite{IvanovaMichaelides2023}  & Number of actions & Number of actions a student performs on an item. \\
\cite{Self2013} & Participation & Homework completion and voluntary practice access. \\
\bottomrule
\end{tabularx}
\end{table}

At the same time, these proxies are limited. Time-based measures are shaped not only by effortful cognition but also by task demands (e.g., reading load, algebraic complexity), interface constraints \cite{baker2008students}, classroom interruptions, and off-task behavior \cite{gurung2025starting}. As a result, the same observed delay can reflect careful deliberation for one student and idling or confusion for another. For this reason, much prior work has used response time primarily at the extremes: unusually fast responses are treated as evidence of rapid guessing or low-effort responding \cite{baker2007gaming,gurung2021examining,kong2007setting}. This use case is valuable, but it leaves open a harder measurement problem central to multi-step problem solving in tutoring systems: whether typical step-to-step timing contains predictive information about \emph{constructive} engagement, once differences in content difficulty and step structure are accounted for. This conjecture is motivated by recent evidence that engagement differences between learners can be highly stable across time \cite{gurung2025starting,huang2023using}.

Our contribution builds on this measurement tradition and on classic psychometric work that models response time alongside accuracy \cite{thissen1983timed}. We estimate \emph{response-time propensities}: difficulty- and context-adjusted, trait-like differences in students' typical response timing during tutoring. The term ``trait-like'' captures typical pacing at the step level; within-person speed can still vary across steps. This makes it more interpretable and allows comparisons of step completions beyond the speed of individual attempts, which can take the form of hints or problem-step solutions in ITSs \cite{koedinger2007exploring}.

\subsection{Effort and Prior Knowledge}

Prior knowledge shapes memory encoding, consolidation, and retrieval processes, enhancing learning when new information is congruent with existing knowledge but hindering learning when it conflicts with learners' prior beliefs \cite{ShingBrod2016Effects}. As a result, prior knowledge often moderates the effectiveness of learning effort: knowledgeable learners tend to learn new information more efficiently within the same domain \cite{WitherbyCarpenter2022RichGetRicher} and can often achieve strong performance with relatively less additional effort \cite{Zambrano2019Effects}. At the same time, this relationship is not monotonic. For certain complex tasks, higher prior knowledge may increase perceived intrinsic cognitive load because experts form more accurate and elaborated problem representations, whereas novices may underestimate task complexity \cite{Endres2022CanPrior}.

Across instructional contexts, a consistent empirical principle is that learning is optimized when the level of effort and instructional support is well matched to learners' prior knowledge \cite{kalyuga2009expertise,koedinger2007exploring}. Too little effort or support can lead to shallow processing, while excessive demands can overwhelm learners and produce unproductive struggle. Importantly, these dynamics suggest that the same observable behavior (e.g., spending more time on a problem-solving step) may reflect qualitatively different processes depending on what the learner already knows. For higher-knowledge students, extended engagement may signal constructive processing and self-explanation; for lower-knowledge students, it may instead indicate confusion, stalled progress, or disengagement.

Despite the centrality of prior knowledge in theories of learning and instructional design, most large-scale log-based analyses treat effort signals as uniformly interpretable across learners \cite{baker2007gaming,gurung2021examining,huang2023using,kovanovic2015does}. This leaves a critical gap: we lack scalable models that capture how the \emph{meaning} of behavioral effort varies with prior proficiency during multi-step problem solving. The present study addresses this gap by modeling response time as a difficulty-adjusted, trait-like propensity and by explicitly testing how students' prior knowledge moderates its relationship with learning efficiency.

\subsection{Effort and Learning Outcomes}

Prior studies have shown that students' time investment is often a useful predictor of academic performance in online learning contexts \cite{Aydogdu2020ANN}. Related work in written comprehension tasks similarly indicates that students who spend more time on a task and produce more extensive responses tend to demonstrate deeper understanding of the material \cite{BratenLatiniHaverkamp2022Engagement}. At the same time, a growing body of evidence suggests that the relationship between time-based engagement and learning outcomes is far from uniform and is contingent on both contextual and learner-specific factors \cite{SCHNEIDER2025102744}.

For example, response time has been found to be positively related to item accuracy. Still, this association is moderated by item cognitive level: longer response times are more strongly associated with higher performance on higher-order cognitive items than on lower-level items \cite{Liu2021ResponseTime}. Similarly, studies of instructional interventions in adaptive learning systems report cases in which interventions successfully increase students' time investment without producing corresponding gains in mastery, indicating that additional time does not necessarily translate into more effective learning \cite{Vanacore2023NonCognitive}. Consistent with this view, research on predicting student performance in MOOCs has shown that extended time on task may sometimes reflect confusion, inefficient strategies, or stalled progress rather than greater productive effort \cite{Lee2021MOOC}. Together, these findings suggest that time-based measures of effort must be interpreted in relation to when, how, and by whom time is invested.

In classroom-based intelligent tutoring system deployments, one particularly salient yet underexplored contextual factor is \emph{when} effort occurs during a class session. Recent work shows that students who delay starting work at the beginning of a session are substantially less likely to persist and learn, even when total available practice time is held constant \cite{gurung2025starting}. Early in a session, students are also more likely to encounter new content or to engage with lower levels of mastery, making effort during this period potentially more diagnostic of productive engagement and self-regulation. In contrast, later in the session, fatigue, task switching, and diminishing motivation may reduce both engagement and the interpretability of time-based behaviors. The present study builds on this insight by treating position within the classroom session as a moderator and examining whether response-time propensities are most predictive of learning efficiency early in practice, when effort regulation is most consequential.

\subsection{Student Trait Modeling}
\label{sec:background:ltt}

Early work in learning analytics distinguishes between \emph{state-based} and \emph{trait-based} explanations of learner behavior. A foundational example is Baker's analysis of gaming the system, which showed that gaming is better explained by situational factors than by stable learner traits \cite{baker2007gaming}. Subsequent work has adopted hybrid approaches, combining state and trait components to model unproductive behaviors more precisely \cite{huang2023using}. More recently, research using tutoring system log data suggests that persistence and engagement behavior in tutoring systems also exhibits strong trait-like stability \cite{Dang2022,gurung2025starting}.

Trait-based perspectives offer an important advantage: when defined relative to task demands, they can remain interpretable across contexts. Prior work shows that engagement depends on the match between learner competence and task difficulty \cite{lynch2019exploring,papouvsek2016impact}, and that difficulty can systematically influence the likelihood of unproductive behaviors \cite{baker2009educational}. Consistent with this view, recent large-scale studies have demonstrated that stable, trait-like patterns of learning behavior can be recovered from log data and can predict engagement and learning outcomes independently of performance \cite{AkhuseyinogluBrusilovsky2021,MirzaeiSahebiBrusilovsky2020}.

Despite these advances, trait modeling in intelligent tutoring systems has focused almost exclusively on unproductive behaviors, such as gaming \cite{huang2023using}, or on overall session time usage \cite{gurung2025starting}. To date, there has been little work on trait-like models of \emph{constructive effort} that explicitly adjust for content and task variation. Addressing this gap is critical, given the large and persistent differences in how much students practice and benefit from adaptive learning systems in real-world classrooms \cite{eames2026computer,koedinger2023astonishing}.

\section{Datasets and Tutoring Software}
\label{subsec:datasets}

\begin{table*}[htpb]
\centering
\caption{Summary of datasets included in the analysis, including their associated research project citation and their associated DataShop (DS) dataset ID for reproducibility.}
\label{tab:datasets}
\small
\begin{tabular}{lcllcrrrrr}
\hline
\textbf{Dataset} & \textbf{DS ID} & \textbf{Citation} & \textbf{Focus} & \textbf{Year} & \textbf{Students} & \textbf{Actions} & \textbf{Steps} & \textbf{Problems} & \textbf{Mins./Student} \\
\hline
Paper vs. Tutor (Tutor Condition) & 5360 & \cite{borchers2023makes}     & Paper vs. Tutor      & 2022 & 96  & 12{,}091 & 6{,}107  & 871     & 30.2 \\
Gamification Study (Cohort 1)            & 5605 & \cite{borchers2025more}      & Gamification & 2023 & 53  & 9{,}635  & 3{,}706  & 1{,}452 & 54.5 \\
Gamification Study (Cohort 2)            & 5606 & \cite{borchers2025more}      & Gamification & 2023 & 96  & 19{,}171 & 8{,}229  & 3{,}554 & 60.1 \\
Drag\&Drop Equation-Solving           & 5610 & \cite{borchers2025more}      & Input Modality & 2023 & 36  & 1{,}973  & 431      & 200     & 29.2 \\
Equation Collaboration Study          & 3332 & \cite{borchers2024combining} & Peer Tutoring        & 2020 & 120 & 10{,}381 & 8{,}276  & 2{,}358 & 39.8 \\
Equation Collaboration Study II            & 5153 & \cite{borchers2024combining} & Peer Tutoring        & 2021 & 198 & 18{,}704 & 12{,}072 & 5{,}145 & 48.7 \\
Equation Collaboration Study III                   & 5604 & \cite{borchers2024combining} & Peer Tutoring        & 2023 & 91  & 6{,}921  & 2{,}274  & 1{,}700 & 35.1 \\
Dynamic Collaborative Transitions          & 5549 & \cite{borchers2024combining} & Peer Tutoring        & 2023 & 106 & 8{,}716  & 3{,}709  & 2{,}250 & 42.7 \\
\hline
\textbf{Total}                     & –   & –                           & –                   & –   & 794 & 87{,}592 & 44{,}804 & 17{,}530 & 43.7 \\
\hline
\end{tabular}
\end{table*}

To demonstrate learning process patterns at scale, we analyzed eight datasets collected between 2020 and 2023 from IRB-approved classroom studies examining student problem solving in intelligent tutoring systems. All studies were conducted in authentic middle- and high-school mathematics classrooms in the U.S. In each study, students were required to use the tutor during scheduled sessions unless parents opted out. We requested access to these datasets via DataShop, an open research data repository for educational data \cite{koedinger2010data}. The approved protocols permitted analysis of anonymized student interaction logs collected during normal instructional activities. Parents and legal guardians were able to opt their child out of the research use of their log data. Across studies, students used the same class of step-based intelligent tutoring systems, targeting equation-solving and linear algebra skills, though individual experiments varied instructional conditions. The log data were standardized as all experiments used the CTAT+TutorShop research infrastructure \cite{aleven2025integrated}. All tutoring systems shared core ITS features, including step-level adaptive guidance \cite{vanlehn2011relative}. Some studies used individualized mastery learning \cite{corbett2000modeling}; others (e.g., gamification, drag-and-drop \cite{borchers2025more}) used skill estimates for display without mastery-based problem selection. Systems estimated skill mastery (e.g., Bayesian Knowledge Tracing \cite{corbett1994knowledge}) and, where applicable, used these estimates to select subsequent problems (see studies linked in Table \ref{tab:datasets} for details on ITS configurations).

All tutoring systems logged student interactions at the level of individual problem-solving steps, including anonymized student identifiers, problem and step identifiers, timestamps, associated knowledge components, and outcome codes (correct, incorrect, hint). A \emph{learning opportunity} is defined as a distinct problem-solving step associated with at least one knowledge component, as is common in tutoring systems research~\cite{borchers2024combining}. Students are assumed to learn from these opportunities through hints and feedback. Accordingly, learning is inferred from step-level performance, where accuracy is measured based on a student's unassisted first attempt at each problem-solving step, consistent with standard practices in ITS research~\cite{koedinger2010data}. All datasets were processed using an open-source preprocessing pipeline that standardized datasets into a shared format.\footnote{\url{https://github.com/conradborchers/response-time-propensities/}}

Table~\ref{tab:datasets} summarizes the datasets included in the analysis, along with their associated publications, study focus, and DataShop identifiers for reproducibility. The datasets contain approximately 30--60 minutes of practice time for a total of 794 students. This corresponds to approximately 1--2 classroom sessions per student. Since all datasets were collected in comparable classroom settings with about 45 minutes of total class session time, we can analyze all data using session-level time slicing \cite{gurung2025starting}, as described next.

\begin{figure}[t]
\centering
\includegraphics[width=\linewidth]{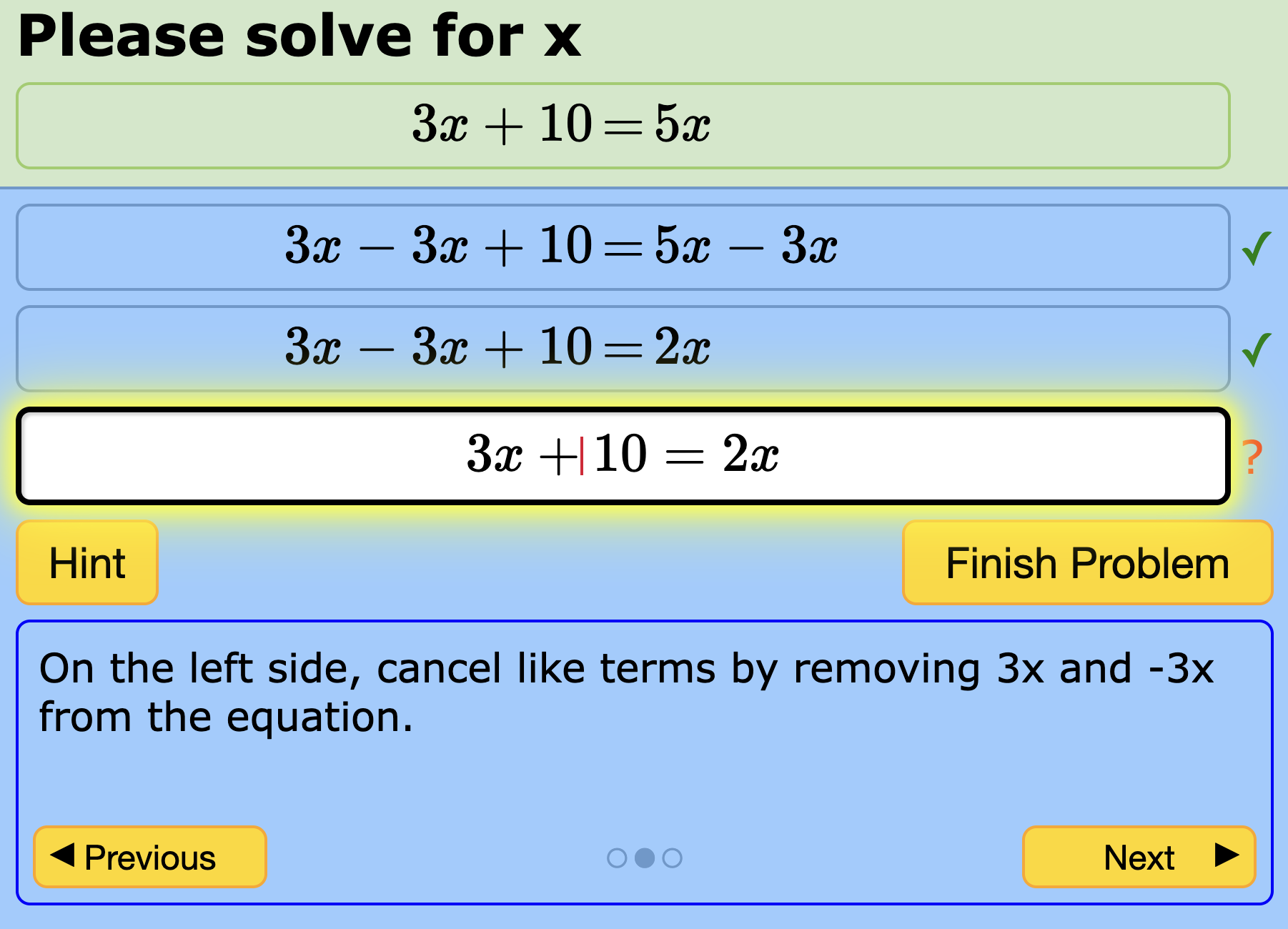}
\caption{The Lynnette intelligent tutoring system interface for equation solving with an example step hint. Equation transformation rows represent steps. While the specific interface design varied by study, this image represents the canonical type-in version of Lynnette and its core features.}
\label{fig:lynnette}
\end{figure}

All studies included data from the tutoring system for equation-solving, based on the open-source Lynnette tutor \cite{long2018exactly}, except for the paper vs. tutor study \cite{borchers2023makes}, which primarily employed comparable systems based on graph interpretation. The Lynnette tutor is shown in Figure \ref{fig:lynnette}. The Lynnette tutor is an intelligent tutoring system designed to support students in solving algebraic equations through step-by-step interaction. Rather than submitting only final answers, students enter intermediate solution steps, which the system evaluates for both algebraic equivalence and progress toward isolating the target variable $x$. When an incorrect or unproductive step is entered, the system provides immediate feedback and allows students to revise their work. In addition to step-level feedback, the Lynnette tutor supports on-demand hint requests, a common feature of intelligent tutoring systems. Hints are delivered in a graduated manner, ranging from general strategic guidance to more specific procedural suggestions. Student interactions with hints, including request timing and frequency, are logged alongside problem-solving actions, enabling analysis of help-seeking behavior and learning processes.

Across students, learners engaged with an average of 7.44 unique skills (SD = 2.28; Median = 7; IQR = 3). Each skill was practiced on average 10.84 times per student (SD = 7.48; Median = 9.48; IQR = 7.73), and students completed 87.61 total skill observations on average (SD = 72.47; Median = 73; IQR = 80), indicating substantial variability in overall practice exposure. This distribution, as well as the required amount of practice to achieve mastery (automatically assessed by the ITS), is comparable to past tutoring system research \cite{koedinger2023astonishing}. The dataset included 32 distinct skills spanning graph construction (e.g., drawing lines, plotting points), evaluating linear functions (e.g., calculating y from x, identifying x from y), interpreting slope and intercepts, identifying independent and dependent variables and their units, and performing algebraic manipulations such as combining like terms, distributing, dividing, and canceling terms.
Further details about the ITS and an evaluation study of its skill models are described in Long et al. \cite{long2018exactly}.

\section{Methods}

Below, we describe our measures and analytical approaches.

\subsection{Response Time and Opportunity Measures}

Response times (RTs) were calculated as the elapsed time between consecutive first attempts on problem-solving steps, capturing the time students spent progressing between steps such as ``subtracting 3 from both sides'' in individually tutored problems. A problem-solving step is a discrete action opportunity that may be associated with one or more knowledge components. We use steps rather than whole items or problems because each step yields a single latency aligned with a specific skill application, which allows us to separate skill-level difficulty from student-level pacing in the hierarchical model. A step can include multiple attempts if students engage in rapid guessing, a common phenomenon in tutoring system research \cite{baker2008students}. RTs were log-transformed to account for skew. To model learning opportunities, each student $\times$ skill sequence was enumerated, producing an opportunity count for each step, in line with standard approaches to modeling learning in tutoring systems \cite{borchers2024combining,koedinger2023astonishing}. Students' classroom sessions were normalized into four global time slices (Q1–Q4) representing the earliest to the latest quartiles of their sessions. A session was defined as the start and end times when students began practicing in their math software.

\subsection{Latent Student and Item Modeling}

To estimate latent differences in student effort and learning while accounting for contextual variation, we fit two complementary mixed-effects models: one modeling step-level response time as a proxy for effort, and the other modeling learning across practice opportunities. In both cases, we adjusted for variation in problem-step difficulty by including random effects at the skill level, consistent with prior work on trait modeling in tutoring systems \cite{huang2023using}. All models were estimated using the \texttt{lme4} package in R \cite{bates2015fitting} and were fit separately within global time slices.

\paragraph{Response-Time Propensity Model.}
To capture stable individual differences in how students allocate time during multi-step problem solving, we modeled log-transformed step-to-step response times using a random-intercepts model:
\begin{equation}
\label{eq:rt}
\mathrm{rt\_log} \sim 1 + (1 \mid \mathrm{student}) + (1 \mid \mathrm{skill}).
\end{equation}
This specification decomposes observed response times into a global mean, a student-level random intercept capturing each learner's typical response timing (response-time propensity), and a skill-level random intercept capturing systematic differences in step difficulty and structure. The resulting student random effects provide a difficulty-adjusted, trait-like measure of students' typical engagement pacing during tutoring, independent of correctness.

\paragraph{Learning Efficiency Model.}
To estimate how efficiently students learn from practice, we modeled step-level correctness as a function of practice opportunity counts using the individualized Additive Factors Model (iAFM) \cite{liu2017towards}:
\begin{equation}
\label{eq:iafm}
\text{correct} \sim \text{opp.} + (1 + \text{opp.} \mid \text{student}) + (1 + \text{opp.} \mid \text{skill}).
\end{equation}
In this model, the fixed effect of opportunity captures average learning over individual practice opportunities, while student-level random slopes represent individual learning rates (learning efficiency differences). The model assumes that, with additional practice opportunities, students become more likely to produce correct first attempts on steps involving a given skill (e.g., repeatedly practicing ``addition''). Skill-level random effects account for differences in baseline difficulty and learnability across skills. The iAFM is a widely used and validated approach for modeling learning in intelligent tutoring systems and has been applied extensively in tutoring system research \cite{borchers2024combining,koedinger2023astonishing,liu2017towards}.

Student-level random effects from both models were extracted and used in subsequent analyses to examine the stability of response-time propensities and their relationship to learning efficiency over time and across learner contexts.

\subsection{Analyses}

We conducted three analyses aligned with our research questions, using student-level parameters extracted from the hierarchical models described above.

\textbf{RQ1 (Stability).} To assess the stability of response-time (RT) propensities and learning-rate estimates over practice, we computed pairwise Pearson correlations of student random-effect estimates across global time slices (Q1–Q4). We report the full cross-slice correlation patterns for both RT intercepts and learning-rate slopes, and we highlight early-to-late stability using the Q1–Q4 correlation.

\textbf{RQ2 (RT and learning associations).} To test whether RT propensity is related to learning efficiency, we computed within-slice Pearson correlations between students' RT intercepts and their learning-rate slopes (opportunity slopes) in each time slice. This provides a descriptive test of whether the RT–learning relationship differs depending on when the behavior occurs in practice.

\textbf{RQ3 (Contextual moderation).} To test whether prior proficiency moderates the RT–learning relationship, we fit a slice-specific linear model predicting each student's learning efficiency from their RT propensity, baseline proficiency, and their interaction:
\[
\mathrm{learning\_rate} \sim \mathrm{RT\_propensity} \times \mathrm{prior\_proficiency}.
\]
We operationalized \textit{prior knowledge} as the student random intercept from the iAFM (i.e., baseline correctness after accounting for skill effects and opportunities), which is a conventional interpretation of this parameter \cite{liu2017towards,koedinger2023astonishing}. RT propensity and learning rate were taken as the student random intercept from the RT model and the student random slope on opportunity from the iAFM, respectively. We report coefficients with 95\% confidence intervals and raw $p$ values, and, for within-slice analyses, control the false discovery rate across the four slices using Benjamini–Hochberg adjustments applied separately for each effect. To aid interpretation, we visualize (i) the interaction coefficient by slice with confidence intervals and (ii) fitted RT–learning relationships at representative proficiency levels (e.g., $\pm1$ SD), alongside binned proficiency groups. All variables were standardized to a mean of 0 and a standard deviation of 1 to enhance interpretation.

\section{Results}

\begin{figure}[ht]
\centering
\includegraphics[width=\linewidth]{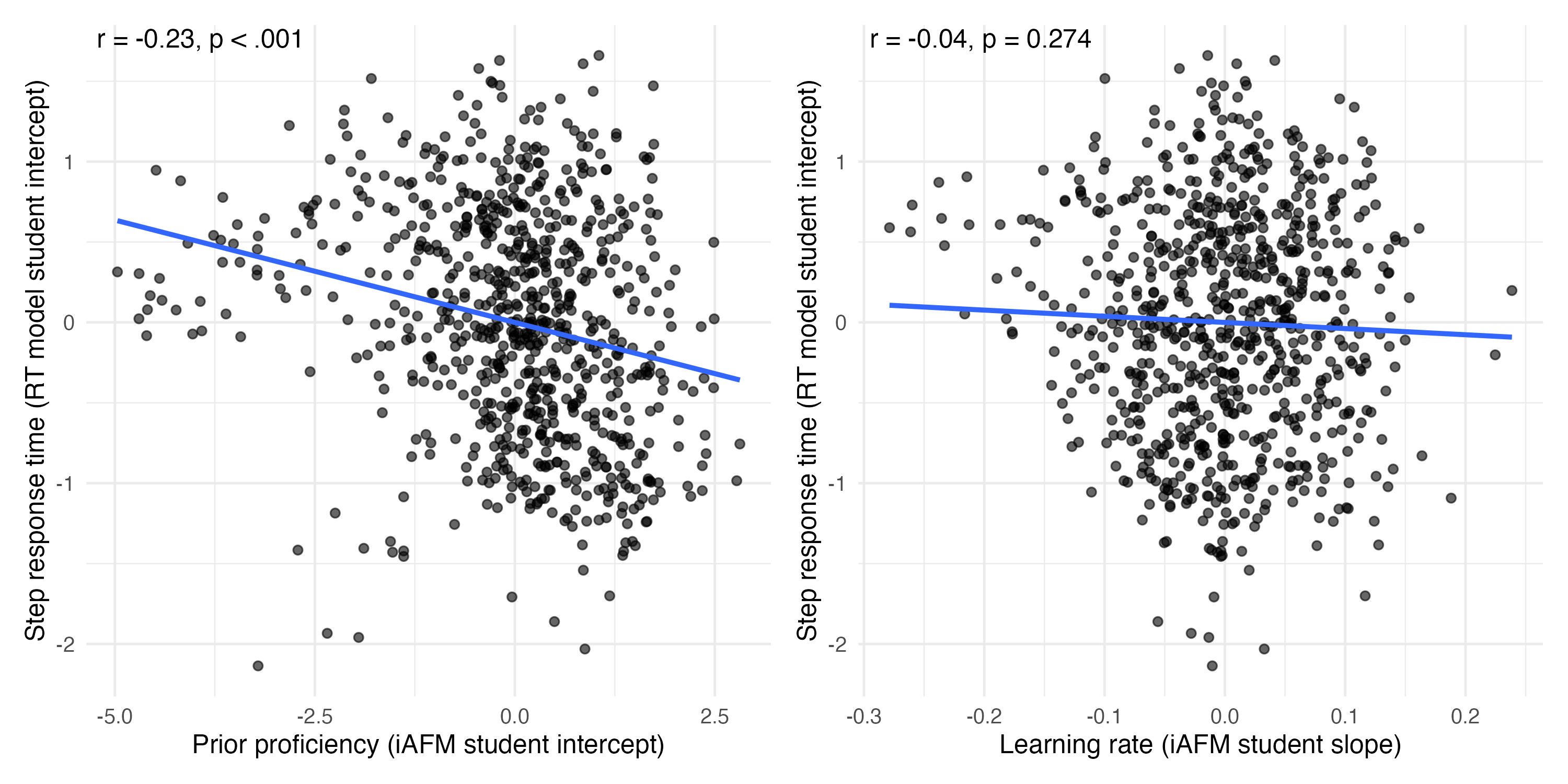}
\caption{Relationship between student learning parameters and response-time (RT) propensity.}
\label{fig:rt_learning_relationship}
\end{figure}

Figure~\ref{fig:rt_learning_relationship} illustrates the relationship between student-level response-time (RT) propensity and learning-related parameters from the iAFM model. The left panel shows a modest negative correlation between students' prior proficiency (iAFM student intercept, Eq.~\ref{eq:iafm}) and RT propensity (RT model student intercept, Eq.~\ref{eq:rt}); the right panel shows the correlation between learning rate (iAFM student slope on opportunity, Eq.~\ref{eq:iafm}) and RT propensity. Higher proficiency is associated with faster step-level responses (left panel); there is no reliable relationship between learning rate and RT propensity across the full practice period (right panel).

\subsection{Construct Stability (RQ1)}

Response-time propensities showed moderate to strong stability across practice, with correlations between early and later slices ranging from $r=.51$ (Q1–Q3) to $r=.55$ (Q1–Q4), and the strongest association observed between adjacent late slices (Q3–Q4; $r=.64$). In contrast, learning rate estimates were less stable, with correlations ranging from $r=.19$ (Q2–Q4, Q3–Q4) to $r=.35$ (Q1–Q2), indicating greater sensitivity to contextual variation across the session. Table~\ref{tab:rt_quartiles} reports descriptive statistics for step-level response time by quartile. Median step RT was roughly stable across quartiles (about 19–20 s).

\begin{table}[ht]
\centering
\caption{Step-level response time (seconds) by session quartile (median [IQR]).}
\label{tab:rt_quartiles}
\small
\begin{tabular}{lcccc}
\toprule
 & Q1 (earliest) & Q2 & Q3 & Q4 (latest) \\
\midrule
RT & 20.3 [41.9] & 18.8 [38.1] & 18.8 [37.1] & 20.0 [41.7] \\
\bottomrule
\end{tabular}
\end{table}

\subsection{Relationship Between RT Propensity and Learning Rate (RQ2)}

\begin{figure}[H]
\centering
\includegraphics[width=\linewidth]{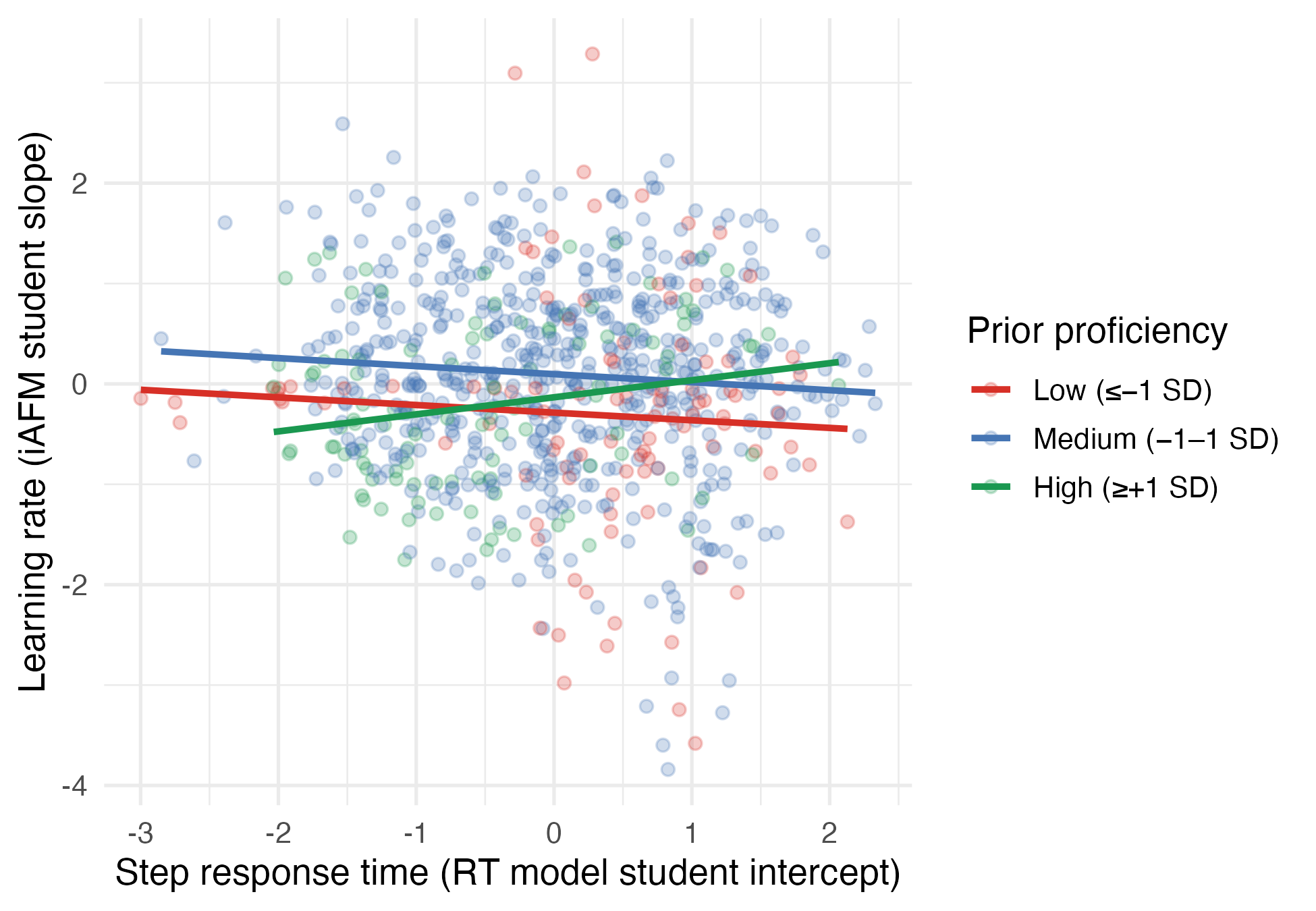}
\caption{Relationship between student-level response-time (RT) propensity and learning rate, moderated by prior proficiency. Points represent students; lines show linear fits by proficiency group.}
\label{fig:rt_learning_interaction}
\end{figure}

Student-level analyses reveal a systematic interaction between response-time (RT) propensity and prior proficiency in predicting learning rate. Prior proficiency is negatively associated with RT propensity (Figure~\ref{fig:rt_learning_relationship}), indicating that more proficient students tend to respond more quickly at the step level. RT propensity shows no reliable main effect on learning rate ($\beta=-0.02$, 95\% CI $[-0.09, 0.05]$, $p=.534$), whereas prior proficiency is positively associated with learning rate ($\beta=0.10$, 95\% CI $[0.02, 0.17]$, $p=.009$). Critically, there is a significant RT propensity $\times$ prior proficiency interaction ($\beta=0.11$, 95\% CI $[0.03, 0.18]$, $p=.004$; Table~\ref{tab:interaction}), indicating that the association between RT propensity and learning rate differs by students' baseline proficiency. As shown in Figure~\ref{fig:rt_learning_interaction}, longer RT propensities are associated with steeper opportunity slopes among higher-proficiency students, are flat among students with intermediate proficiency, and are weakly negative among lower-proficiency students.

\begin{table}[H]
\centering
\caption{Predictors of learning rate (opportunity slope) from RT and learning intercepts.}
\label{tab:interaction}
\resizebox{\linewidth}{!}{
\begin{tabular}{lccc}
\hline
Predictor & Estimate & 95\% CI & p \\
\hline
Intercept & 0.02 & [-0.05, 0.10] & 0.499 \\
RT intercept & -0.02 & [-0.09, 0.05] & 0.534 \\
Learning intercept & 0.10 & [0.02, 0.17] & 0.009 \\
RT intercept $\times$ Learning intercept & 0.11 & [0.03, 0.18] & 0.004 \\
\hline
\end{tabular}
}
\end{table}

Visual inspection of Figure \ref{fig:rt_learning_interaction} suggested a small cluster of slow-responding students with low learning rates. Because observations in extreme regions of the predictor–outcome space can exert disproportionate influence on interaction estimates in linear models, we conducted a sensitivity analysis. We flagged potentially influential observations (Cook's D > 4/N, leverage > 2p/N, or |studentized residual| > 3) and re-estimated the model excluding these cases (98/787, 12.5\%). The RT propensity--prior proficiency interaction remained significant ($\beta$ = 0.147, p = .014).

\subsection{Time Slice Analyses of Context (RQ3)}

To examine whether the relationship between RT propensity and learning varies over the course of a classroom session, we analyzed student parameters within four global time slices (Q1–Q4). The correlation between RT propensity and learning rate is strongest early in the session and attenuates over time. In the earliest slice (Q1), RT propensity is modestly but significantly negatively correlated with learning rate ($r=-0.12$, $p<.001$), whereas correlations in later slices are smaller and not statistically reliable (Q2: $r=-0.05$, $p=.145$; Q3: $r=0.04$, $p=.296$; Q4: $r=-0.03$, $p=.375$).

Regression analyses that account for prior proficiency (Table~\ref{tab:slice_interaction}) reveal a similar pattern. The interaction between RT propensity and prior proficiency is significant in the earliest slice (Q1; $\beta=0.12$, 95\% CI $[0.03, 0.20]$, adjusted $p<.05$) and marginally significant in the latest slice (Q4; $\beta=0.09$, 95\% CI $[0.01, 0.17]$, adjusted $p<.10$), but not in the middle slices (Q2 and Q3). Across all slices, RT propensity shows no reliable main effect on learning rate, whereas the effect of prior proficiency varies in both magnitude and direction over time. Together, these results indicate that the association between response timing and learning efficiency is strongest early in practice and becomes less systematic as the session progresses.

\begin{table}[htpb]
\centering
\caption{Time-slice–specific regression coefficients from linear models predicting student learning rate as a function of response-time (RT) propensity, prior proficiency, and their interaction.}
\label{tab:slice_interaction}
\small
\resizebox{\linewidth}{!}{
\begin{tabular}{llcccc}
\hline
Slice & Effect & Estimate & 95\% CI & $p$ & Sig. \\
\hline
Q1 & Intercept & 0.034 & $[-0.039, 0.108]$ & 0.363 &  \\
Q2            & Intercept & -0.005 & $[-0.078, 0.068]$ & 0.893 &  \\
Q3            & Intercept & 0.014 & $[-0.057, 0.085]$ & 0.701 &  \\
Q4   & Intercept & 0.016 & $[-0.055, 0.087]$ & 0.653 &  \\
\hline
Q1 & Prior proficiency & 0.243 & $[0.170, 0.316]$ & $<.001$ & *** \\
Q2            & Prior proficiency & -0.084 & $[-0.157, -0.010]$ & 0.025 & * \\
Q3            & Prior proficiency & -0.237 & $[-0.308, -0.167]$ & $<.001$ & *** \\
Q4    & Prior proficiency & 0.110 & $[0.039, 0.180]$ & 0.002 & ** \\
\hline
Q1 & RT propensity & -0.002 & $[-0.075, 0.070]$ & 0.951 &  \\
Q2            & RT propensity & 0.021 & $[-0.052, 0.094]$ & 0.568 &  \\
Q3            & RT propensity & -0.018 & $[-0.089, 0.053]$ & 0.619 &  \\
Q4  & RT propensity & -0.038 & $[-0.109, 0.033]$ & 0.288 &  \\
\hline
Q1 & RT $\times$ proficiency & 0.116 & $[0.032, 0.199]$ & 0.007 & * \\
Q2            & RT $\times$ proficiency & -0.018 & $[-0.091, 0.054]$ & 0.622 &  \\
Q3            & RT $\times$ proficiency & 0.065 & $[-0.020, 0.150]$ & 0.135 &  \\
Q4   & RT $\times$ proficiency & 0.092 & $[0.011, 0.173]$ & 0.026 & . \\
\hline
\end{tabular}
}
\footnotesize
\textit{Note.} Raw $p$ values are reported. Significance markers are based on
Benjamini–Hochberg adjusted $p$ values computed for each effect
(*** $p<.001$, ** $p<.01$, * $p<.05$, . $p<.10$).
\end{table}

\section{Discussion}

We asked (RQ1) how reliable response-time propensities are during multi-step problem solving; (RQ2) to what extent they predict learning efficiency; and (RQ3) how contextual factors (early vs. late practice, prior proficiency) influence the RT–learning relationship. This section interprets the results in light of these questions.

Across eight classroom deployments spanning six schools, we modeled response time as a difficulty-adjusted propensity, separating stable student differences from systematic variation due to knowledge components. The results support two complementary conclusions. First, response-time propensities are reliable enough to sustain a trait-like interpretation within classroom practice, showing moderate to strong stability across session slices (RQ1). Second, response time is not uniformly ``good'' or ``bad'' for learning; its association with learning efficiency depends on who the learner is and when in the session the behavior is observed (RQ2–RQ3). In particular, slower propensities predict greater learning efficiency among higher-proficiency students, whereas among lower-proficiency students, they are weakly or even negatively associated. This pattern is most diagnostic early in practice and becomes less systematic as the class period progresses.

\subsection{Response-Time Propensities as a Context-Sensitive Effort Signal at Scale}

A central contribution of this work is to show how step-to-step response time can be made interpretable at scale by reframing it as a propensity instead of a raw delay. Response time is ubiquitous in tutoring logs and has long been used as a pragmatic indicator of behavioral engagement in computer-based learning and assessment \cite{WiseKong2005}. At the same time, prior research has emphasized that time-based measures are intrinsically ambiguous, because they conflate cognitive processing with task demands, interface constraints, and off-task interruptions \cite{kovanovic2015does,kovanovic2015penetrating}. This ambiguity has led much of the ITS literature to rely primarily on response time at its extremes, most notably as a marker of rapid guessing or other low-effort responding \cite{baker2007gaming,gurung2021examining,kong2007setting}. Our results support caution while also offering a constructive path forward. By modeling log response times with student and knowledge-component random effects, we estimate a difficulty-adjusted student parameter that isolates each learner's typical pacing during multi-step work. The resulting response-time propensities are stable enough within classroom practice to support an individual-differences interpretation, addressing RQ1 and aligning with recent evidence that some engagement behaviors exhibit trait-like consistency when defined relative to task demands \cite{huang2023using,gurung2025starting}.

This shift from raw time to propensity matters because it changes what response time can be used for in \textbf{learning at scale}. The field has robust methods for tracking knowledge and instructional effectiveness from correctness trajectories \cite{corbett1994knowledge,koedinger2023astonishing,liu2017towards}, yet large classroom deployments continue to show that many students do not engage long enough or deeply enough to benefit from otherwise effective systems \cite{eames2026computer,holt20245,koedinger2015learning}. Response-time propensities contribute an additional axis of learner modeling (and potentially novel forms of adaptivity and classroom interventions) that is not reducible to correctness alone: they capture how students distribute time across steps, a behavioral footprint of regulation that is largely invisible in performance-only models. In this sense, the propensities are best understood as a process construct that complements knowledge estimates. The stability we observe also helps clarify why session-level measures, such as time on task, can be noisy proxies for effort. When pacing is modeled relative to skills, it becomes possible to compare students in a way that is less entangled with which problems they happened to receive or how difficult those problems were \cite{kovanovic2015penetrating}.

\subsection{Contextual Variations in Response Times}

The second implication, addressing RQ2, is that the meaning of response timing is conditional on learner and context, not uniform. Global associations between timing and learning are sometimes weak \cite{kovanovic2015does}. Our findings help explain why. Response-time propensities relate to learning efficiency primarily through moderation by prior proficiency. More proficient students tend to respond more quickly overall. This association is consistent with a fluency account \cite{wang2020using}: as students enter practice with stronger procedural knowledge, they may execute steps more automatically and with less deliberation, yielding shorter latencies without implying low effort.
Still, among those students, slower propensities are associated with greater learning efficiency, consistent with constructive processing, monitoring, or self-explanation, which convert additional time into learning gains \cite{kalyuga2009expertise,koedinger2007exploring}. For less proficient students, slower propensities are flat or weakly negative with respect to learning, consistent with the idea that longer delays are more likely to reflect stalled progress, inefficient strategy search, or disengagement that is difficult to distinguish from deliberation in log data alone, since curricular context can shift over time \cite{vanacore2025downshifting}. This interaction is important for interpretation and for the design of learning at scale. It implies that timing-based signals cannot be used as one-size-fits-all indicators of effort quality, because the same observable pacing can reflect qualitatively different processes depending on what the learner already knows \cite{kalyuga2009expertise}. It also implies that naive interventions that treat slow responding as disengagement risk disrupting productive work for higher-proficiency students, while failing to provide the targeted support that lower-proficiency students may need when delays signal unproductive struggle.

Finally, the propensity framing provides a principled way to connect temporal process traces to the broader trait-versus-state debate in learning analytics. Classic work on gaming the system emphasized situational explanations and cautioned against trait interpretations \cite{baker2007gaming}. Subsequent models have shown that hybrid approaches, which account for both stable tendencies and contextual differences, can improve behavioral inference \cite{huang2023using}. Our results are compatible with that trajectory. Response-time propensities exhibit trait-like stability, yet their relationship to learning depends on learner context, and, as shown in the slice analyses, their predictive value concentrates in particular moments of practice. This supports a view in which timing features can inform learner modeling when they are defined relative to task structure and interpreted in conjunction with proficiency. For learning at scale, the broader implication is that temporal signals are most useful not when they provide a single global ranking of effort, but when they enable context-aware decisions about when and for whom additional support is warranted \cite{koedinger2023astonishing,koedinger2015learning}.

\subsection{Response Time Signals Vary Across the Class Session}

Another key finding is that response-time propensities are not equally informative throughout a classroom session. When we index student work by relative position within their practice time (typically 20--40 minutes per practice session), the association between timing and learning efficiency is most structured in the earliest slice and becomes weaker and less systematic later. This is consistent with the broader claim that engagement traces in tutoring logs are shaped not only by the immediate problem-solving demands of a step, but also by session-level dynamics that influence how students enter, sustain, and terminate work \cite{gurung2025starting}, aligning with broader theoretical models of effort regulation \cite{locke2019development}. In other words, a given latency can be interpreted differently depending on whether it occurs during the initial transition into practice or after students have already accumulated learning gains while practicing a given content unit.

When students first begin using the tutor, they are not simply solving algebra steps; they are also initiating and stabilizing a work routine. We use four quartiles to balance granularity with sufficient data per slice; the earliest slice is most predictive and is the focus of interpretation. The self-regulation literature emphasizes that task initiation and effort mobilization are foundational processes that precede sustained engagement \cite{locke2019development}. In classroom-scale deployments, delays in beginning practice are strongly predictive of reduced persistence and weaker outcomes even when total available time is comparable \cite{gurung2025starting}. Our slice results extend this perspective into the within-problem timescale. Early-session response times appear to capture meaningful differences in how students allocate time and attention at the moment when effort regulation is most consequential. This is the period when students encounter new content, calibrate their expectations for difficulty, and decide whether to invest in sense-making or adopt shallow strategies. The moderation by proficiency suggests that early delays can reflect qualitatively different processes: for higher-proficiency students, a slower early tendency is more consistent with deliberate planning and constructive processing that yield greater learning efficiency. In contrast, for lower-proficiency students, early extended latencies are less likely to translate into learning, plausibly reflecting difficulty getting started productively, particularly when content difficulty is high relative to proficiency \cite{kalyuga2009expertise,koedinger2007exploring}.

Later in the session, the predictive validity of timing declines. Step-level response time increases from Q3 to Q4 (Table~\ref{tab:rt_quartiles}), even though content mastery typically rises with practice. The combination of longer latencies and higher mastery late in the session is notable and suggests that end-of-session delays may reflect factors other than productive effort (e.g., fatigue, task switching, or strategic stopping). Related work documents declines in persistence and engagement consistent with depletion or fatigue \cite{Dang2022,dang2020ebb}, and that delays in \emph{starting} practice strongly predict reduced persistence and weaker outcomes \cite{gurung2025starting}. Early-session timing is, therefore, most diagnostic of productive engagement. Under such conditions, later in the session, response times may become less diagnostic, as observed delays may reflect heterogeneous causes, and idiosyncratic end-of-session behaviors could dilute the signal.

\subsection{Implications for Learning at Scale}

These findings suggest that response time can inform adaptive support for learning at scale only when interpreted relative to learner proficiency and instructional context. Simple heuristics that equate fast responding with efficiency or slow responding with disengagement risk misclassifying behavior, particularly for higher-proficiency students whose slower pacing may reflect deliberate, constructive processing, not low effort \cite{kalyuga2009expertise,koedinger2007exploring}. Conversely, extended latencies among lower-proficiency students, especially early in practice, appear less likely to translate into learning and may signal emerging difficulties with task initiation or persistence.

The temporal localization of the effect further clarifies when timing data are most useful. Response-time propensities are most predictive early in a session, when effort regulation is still forming and when differences in engagement have disproportionate downstream consequences. This aligns with prior evidence that early delays in starting practice strongly predict reduced persistence and weaker learning outcomes in large-scale deployments \cite{gurung2025starting}. Taken together, these results suggest that timing-based signals are best used to identify early risk and guide context-sensitive interventions, more so than as global indicators of effort quality.

More broadly, this work supports a process-oriented view of learner modeling in learning at scale. Correctness-based models of learner performance remain essential for estimating knowledge and guiding instructional refinement \cite{koedinger2023astonishing,koedinger2012knowledge}. Still, they do not explain why many students fail to benefit from effective systems \cite{koedinger2015learning,eames2026computer}. Response-time propensities offer a scalable way to incorporate temporal process information into learner models, provided they inform when and for whom adaptive support is most likely to be instructionally meaningful.

\subsection{Limitations and Future Research}

Several limitations point to directions for future work. Most importantly, response time remains an inherently ambiguous behavioral trace. Longer delays can reflect productive deliberation, but they may also capture classroom interruptions, off-task behavior, or time spent reading that cannot be distinguished in log data alone \cite{kovanovic2015penetrating}. Future research may strengthen construct validity by triangulating response-time propensities with complementary indicators of engagement, including observational measures such as those derived from BROMP \cite{ocumpaugh2015baker} or richer process traces. Richer diagnosis of engagement could also be achieved by considering repeated attempts at the same problem-solving steps, such as rapid guessing or reaction times to feedback. Modeling attempt-level dynamics alongside step-level propensities is a direction for future work. 

A second limitation is that the analyses are correlational. The observed moderation by prior proficiency and practice timing is consistent with theoretical accounts of productive struggle and effort regulation, but causal interpretations are not warranted. An important next step is to test whether using early-session timing patterns to trigger adaptive supports improves persistence and learning, for example, through experimental or quasi-experimental designs embedded in tutoring systems. Such studies, conducted over longer intervention periods, could also speak to the long-term stability of response-time propensities and their ability to capture stable student traits (for a discussion, see \cite{borchers2026toward}). Relatedly, incorporating session-level effects in future work could help disentangle stable pacing tendencies from session-specific variation, enhancing estimation of longer-term individual differences \cite{gurung2025starting}.

Finally, the results are bounded by the instructional context studied. All datasets come from step-based mathematics tutors in U.S. classrooms, where interaction cycles are short, and the amount of text to read per step is limited. Response-time propensities may function differently in domains with heavier reading load, open-ended responses, or different classroom norms \cite{baker2019challenges,almoubayyed2023instruction}. Future work should therefore examine whether response-time propensities show comparable stability and predictive value across domains, populations, and system designs before they are used to guide adaptive decisions at scale.

\section{Conclusion}

Step-to-step response time is a widely available trace in adaptive learning systems. Yet, it is often treated as too ambiguous to support general inferences about student effort. This paper shows that the ambiguity is not inherent to timing data itself, but rather to how it is modeled. By reframing step latencies as \emph{response-time propensities} and estimating them with hierarchical models that account for knowledge-component differences, we recover an interpretable, student-level parameter that captures how learners typically pace their work during multi-step problem solving. Across eight classroom deployments of algebra tutoring systems, these propensities display moderate to strong stability within practice sessions, supporting their use as an individual-differences construct that complements performance-based learner models.

At the same time, the results show that response timing cannot be read as a uniform indicator of productive effort. The association between response-time propensity and learning efficiency depends on learner context. For higher-proficiency students, slower propensities are associated with greater learning efficiency, consistent with deliberate, constructive processing that converts time into learning. For lower-proficiency students, slower propensities are weakly or negatively related, consistent with delays that more often reflect stalled progress, inefficient search, or off-task time. This conditional relationship helps explain why global correlations between timing and learning are often small or inconsistent: response time carries meaningful signals that are expressed through interactions with proficiency and context, rather than a single main effect.

The timing signal is also temporally localized within classroom practice. The relationship between propensities and learning is most structured early in a session and becomes less systematic later, when fatigue, interruptions, and strategic stopping behaviors are more likely to introduce heterogeneity into observed delays. This pattern identifies an actionable window in which temporal process data is most diagnostic of how students are engaging with instruction and when early shifts toward disengagement or low persistence may first become detectable.

Taken together, these findings clarify how temporal traces can contribute to learning at scale. Correctness trajectories remain essential for estimating knowledge and guiding instruction. Still, they do not fully explain why many students fail to benefit from effective systems due to differences in persistence, a growing concern in the field. Response-time propensities provide a practical way to incorporate process information into learner models in a manner that is scalable, difficulty-adjusted, and sensitive to learner proficiency and session dynamics. The broader implication is that timing data can support adaptive decisions when interpreted as context-dependent evidence of engagement and regulation, rather than as a direct proxy for effort. Future work can build on this foundation by validating propensities against complementary engagement measures and by testing whether interventions triggered by early-session timing patterns improve persistence and learning in authentic classroom use.

\bibliographystyle{ACM-Reference-Format}
\bibliography{main}

\end{document}